\documentclass[fleqn,final,5p,times,twocolumn,sort&compress,hyperref]{elsarticle}

\usepackage{amssymb}
\usepackage{amsmath}
\usepackage{graphicx}
\usepackage{dcolumn}
\usepackage{bm}
\usepackage{epsfig}
\usepackage{amssymb}
\usepackage{color}
\usepackage{datetime}
\usepackage{wasysym}
\usepackage[mathscr]{euscript}
\usepackage{hyperref}
\usepackage{slashed}
\usepackage{cancel}
\usepackage{ulem}
\usepackage{microtype}
\usepackage[utf8]{inputenc} 

\newcommand{\bal}{\begin{align}}
\newcommand{\eal}{\end{align}}
\newcommand{\beq}{\begin{eqnarray}}
\newcommand{\eeq}{\end{eqnarray}}

\newcommand{\nn}{\nonumber \\}
\newcommand{\es}{& = &}
\newcommand{\rs}{\, = \,}

\newcommand{\cM}{ {\cal M} }
\newcommand{\cH}{ {\cal H} }

\newcommand{\cF}{ {\cal F} }
\newcommand{\cG}{ {\cal G} }

\newcommand{\cQ}{ {\cal Q} }

\newcommand{\cU}{ {\cal U} }
\newcommand{\cL}{ {\cal L} }
\newcommand{\cP}{ {\cal P} }
\newcommand{\cR}{ {\cal R} }

\newcommand{ \rZ  }{ \reflectbox{Z} }
\newcommand{ \trZ }{ {\text{\reflectbox{Z}}} }

\newcommand{\black}{\color[rgb]{0,0,0}}

\newcommand{\h}{ \frac{1}{2} }

\begin{document}

\begin{frontmatter}



\title{ Renormalized quark-antiquark Hamiltonian induced by a gluon mass ansatz in heavy-flavor QCD }


\author[Warsaw,Yale]{ Stanis{\l}aw D. G{\l}azek }
\author[ECTstar]{     Mar\'ia G\'omez-Rocha }
\author[IIT]{         Jai More }
\author[Warsaw]{      Kamil Serafin }

\address[Warsaw]{  Institute of Theoretical Physics,
                  Faculty of Physics, 
                  University of Warsaw,
                  Pasteura 5, 02-093 Warsaw, Poland  }         

\address[Yale]{  Department of Physics,
                 Yale University,
                 217 Prospect Street,
                 New Haven,
                 Connecticut 06511, USA}  

\address[ECTstar]{  European Centre for Theoretical Studies  
                    in Nuclear Physics and Related Areas
                    (ECT*), Villa Tambosi, 
			        38123 Villazzano (Trento), Italy  }
               
\address[IIT]{  Department of Physics,
                Indian Institute of Technology Bombay,
                Powai, Mumbai 400076, India }

\begin{abstract}

In response to the growing need for theoretical tools that can be used 
in QCD to describe and understand the dynamics of gluons in hadrons 
in the Minkowski space-time, the renormalization group procedure for 
effective particles (RGPEP) is shown in the simplest available context of 
heavy quarkonia to exhibit a welcome degree of universality in the first 
approximation it yields once one assumes that beyond perturbation theory 
gluons obtain effective mass. Namely, in the second-order terms, the 
Coulomb potential with Breit-Fermi spin couplings in the effective 
quark-antiquark component of a heavy quarkonium, is corrected in 
one-flavor QCD by a spin-independent harmonic oscillator term that 
does not depend on the assumed effective gluon mass or the choice 
of the RGPEP generator. The new generator we use here is much simpler 
than the ones used before and has the advantage of being suitable for 
studies of the effective gluon dynamics at higher orders than the second 
and beyond the perturbative expansion.

\end{abstract}


\end{frontmatter}



\section{Introduction}

The growing need for understanding dynamics of gluons in QCD, comprehensively documented
in~\cite{EIC}, revives interest in Hamiltonian methods for describing heavy quarkonia in terms 
of wave functions in the Fock space of virtual quanta, where gluons are likely to behave in a 
relatively simple way because they are permanently coupled to the massive quarks that move 
slowly with respect to each other.
 Phenomenology and theory of heavy quarkonia rapidly develops, as is illustrated by many examples~\cite{Brambilla,MatsuiSatz1986,Shan:2017uts,Andronic:2015wma,
Rapp:2008tf,Aaij:2015awa,SzczurekLHC,SzczurekJPsi,SzczurekPANDA,ChineseQuarkonia,Karliner} . The key feature regarding gluons that requires understanding is how they acquire the effective mass, so that the hadron mass spectra ~\cite{PDG2016} do not exhibit any small excitations such as the atomic spectra do due to massless photons. The gluon-mass generation is a subject of research from early on using continuum Dyson-Schwinger equations~\cite{Cornwall:1981zr,CornwallSoni1983,Cornwall:2015lna}
and it is addressed in lattice studies \cite{Bernard:1981pg,Bernard:1982my,Mandula:1987rh,Oliveira:2010xc} because its implications for theory of strong interactions would be broad, including the issue of saturation in dense gluon systems beyond a single hadron setting \cite{Blaizot:2016qgz}. In the canonical formulation of QCD in the front form  (FF) of dynamics in the Minkowski space-time \cite{Dirac1949}, the need for understanding implications of an effective gluon mass is stressed in general in \cite{Wilsonetal} and in the context of heavy quarkonia in \cite{BrisudovaPerryWilson,BrisudovaMassiveGluons}. Theoretical studies of heavy quarkonia may 
thus be expected to increasingly focus on the Hamiltonian dynamics of their gluonic content,  {\it cf.}~\cite{AmesQQHolography,AmesQQLF}.
This article 
presents a first step in a program of systematic studies of dynamics of scale-dependent 
gluons in heavy quarkonia, starting with the simplified case of canonical FF formulation of QCD 
with quarks of just one heavy flavor and assuming that the effective gluons which correspond to 
momentum scale of quark binding mechanism in quarkonia have mass. 

We calculate a renormalized, scale-dependent Hamiltonian for quarkonium constituents 
in the Fock space using a new formulation of the renormalization group procedure for
effective particles (RGPEP) in quantum field theory, see below. In this new formulation, 
the key RGPEP element called its generator does not depend on the derivative of the 
renormalized Hamiltonian with respect to the scale parameter, in contrast to the earlier 
versions of the RGPEP, in which such dependence was involved. The generator dependence 
on the derivative of the Hamiltonian made a calculation of the latter difficult and stalled 
the development. With the derivative issue resolved, the improved method is now prepared
for trial applications and further development. The new generator has been already verified 
to work beyond perturbative expansion in simple theories~\cite{scalarmassmixng, fermionmassmixing} 
and it has passed the test of producing asymptotic freedom in the renormalized FF 
Hamiltonian of QCD~\cite{AF}. Here we present the new-generator 
calculation of a renormalized Hamiltonian in one-flavor theory including terms up to second 
order in powers of the strong-interaction coupling constant at the scale of heavy quarkonia. 

The renormalized Hamiltonian at a suitable scale is subsequently applied to formulate the 
eigenvalue problem for a quarkonium in the Fock space basis that is constructed using 
the creation operators for effective particles of the corresponding size. We make a drastically 
simplifying assumption that the non-Abelian and non-perturbative effects due to components 
with more effective gluons than one, can be mimicked by a gluon-mass ansatz introduced in the
dynamics of the component with only one effective gluon 
and a quark-antiquark 
pair. Namely, the components with more than one effective gluon are dropped at the price of 
introducing the ansatz. We then express the quark-antiquark-gluon component in terms of 
the quark-antiquark component and calculate the effective Hamiltonian in the quark-antiquark 
sector. As a result, we establish the universality of effective interaction one obtains in the 
quark-antiquark component: the new RGPEP generator produces the same Coulomb term 
with Breit-Fermi spin-dependent factors and the same spin-independent harmonic oscillator 
term that were previously obtained with the old generator~\cite{QQ1,QQ2}. The oscillator 
term is 
independent of the value of the mass ansatz. 

The new effective term respects rotational symmetry in the quark-antiquark relative 
three-momentum space, in which the eigenvalue problem for quarkonium two-body 
component resembles a non-relativistic Shr\"odinger equation with a potential, except 
that the eigenvalue is the quarkonium mass squared instead of its energy. The effective 
eigenvalue equation is invariant with respect to the FF kinematical Lorentz transformations. 
We also note that the relative momentum variables we use for heavy quarkonia resemble 
the variables used in the light-front holographic approach to the phenomenology of light 
hadrons, based on the AdS/QCD duality ideas~\cite{holography}. The holographic 
potential is also of the harmonic oscillator form, but its frequency is different, which can  
be associated with much smaller mass of light quarks than $\Lambda_{QCD}$ while 
the heavy quark mass is much greater. 

Section~\ref{sec:ansatz} explains the preliminary nature of the gluon mass 
ansatz that is used to finesse the effective quark-antiquark interaction.  
Section~\ref{secRGPEP} presents the RGPEP in application to one-flavor QCD. 
The renormalized eigenvalue equation for the entire quarkonium state and 
the effective Hamiltonian for its quark-antiquark component are introduced in 
Sec.~\ref{secEVeq}. Section~\ref{secHO} shows how the effective harmonic 
oscillator potential emerges in the non-relativistic limit, using holographic 
quark-antiquark relative momentum variables. Section~\ref{secConclusions} 
concludes the article with comments on how our results prepare ground for 
renormalized Hamiltonian studies of gluon dynamics in heavy quarkonia.

\section{ The initial need for gluon mass ansatz }
\label{sec:ansatz}

The difficulty of describing bound states in the Fock space 
is that one needs to deal with an {\it a priori} infinite number 
of components. In the case of quarkonium, the bound state
\begin{eqnarray}
	|\psi\rangle \es |Q\bar Q\rangle + |Q\bar Q G\rangle + |Q\bar Q GG \rangle + \dots \  ,
\label{QbarQ}
\end{eqnarray}
is built from canonical quanta of quark and gluon fields. To deal with 
this issue, the RGPEP uses the concept of effective particles. They 
are characterized by an effective size $s$ and are related to the bare,
point-like particles of canonical theory by means of an operator 
transformation. The idea is that for description of observables 
characterized by the momentum scale $\lambda=1/s$, 
one can use the renormalized Hamiltonian $H_s$, 
so that
the number of relevant Fock components in Eq.~(\ref{QbarQ}) 
is sufficiently small for carrying out computations, 
except that $Q$, $\bar Q$ and $G$ are 
replaced by $Q_s$, $\bar Q_s$ and $G_s$, respectively. Thus, infinitely 
many Fock components 
for effective particles 
can be neglected. The bound-state problem becomes so greatly 
simplified that one can attempt to seek solutions to the eigenvalue 
equation numerically.

The above strategy may work when the RGPEP equation for $H_s$
is solved exactly. When one uses expansions of $H_s$ in a series of 
powers of effective coupling constant, non-perturbative effects in
$H_s$ itself are not included. The eigenvalue problem for such $H_s$
still couples the Fock components made of effective particles in 
significant ways. Initially, one cannot be certain that the Fock 
components with more than one effective gluon can be dropped
from the eigenvalue problem with no consequence. Since power counting 
allows a mass term for gluons, perturbative RGPEP calculations imply a 
need for a gluon mass counterterm in the canonical Hamiltonian of 
one-flavor QCD, whose finite part is not known, and the phenomenology 
of hadrons appears to exclude massless effective gluons, one is in need 
to assume that neglecting components with more than one effective gluon 
of size $s$ in heavy quarkonia of characteristic physical size $s$ may 
only be 
admissible if one allows for appearance of the effective mass term 
for the gluon in the $Q_s \bar Q_s G_s$ component, presumably produced
by the non-Abelian interaction effects descendant from higher components, 
absent in Abelian theories. 

In the variety of terms that are conceivable in renormalized Hamiltonians 
using FF power counting~\cite{Wilsonetal},  rotational symmetry is
not explicitly respected. 
Any ansatz
one introduces to hypothetically account for the omitted terms, must 
satisfy the condition that the resulting effective eigenvalue problem 
produces the mass spectrum that exhibits degeneracy implied by rotational 
symmetry. Making an ansatz for the gluon mass term, rather than any other 
potentially allowed term, turns out to lead to the effective oscillator correction 
to the Coulomb terms that satisfies the rotational-symmetry condition already 
at the level of effective Hamiltonian itself, 
exhibiting the symmetry in 
relative momentum variables. It is unlikely that ansatz terms for quantities 
other than mass can easily produce such a universal result. Moreover, the 
mass ansatz turns out to possess the unique feature that the oscillator potential 
it leads to is independent of the actual value of the ansatz mass, provided it 
is not too small. 

When the RGPEP calculation of renormalized Hamiltonian $H_s$ is
extended to higher orders than second,
and when 
the unknowns of the effective gluon mass are relegated to the Fock 
sectors with more than one gluon, the preliminary gluon-mass ansatz 
in the dynamics of $Q_s \bar Q_s G_s$ component will be replaced 
by true QCD terms. However, since the complexity of renormalized 
Hamiltonian operators requires approximations, and since their spectra 
are calculated numerically making further simplifications, 
a purely mathematical 
approach to identification of all important terms may turn out hard to 
execute. Fortunately, the spectra of Hamiltonians for heavy quarkonia 
in the Fock space representation with effective gluons can also be studied 
in comparison with experimental data. Assuming that the canonical 
theory one starts from is right, one can use the data in identifying scale 
dependence of various terms allowed by the power counting. The mass 
ansatz seems to be the simplest admissible term to falsify. The universality 
of the quarkonium effective Hamiltonian described here prepares the 
ground for required theoretical studies in parallel with quarkonium 
phenomenology.

\section{ RGPEP as a tool for solving bound states in QCD }
\label{secRGPEP}

Starting from the one-flavor QCD Lagrangian, we derive the renormalized
Hamiltonian $H_s$ which is applied to heavy quarkonia using the effective
particle basis in the Fock space. The eigenvalue problem for $H_s$
is then reduced to the effective quark-antiquark Hamiltonian $H_{s \, \rm eff}$
for the quarkonium $Q_s \bar Q_s$ component. A brief sketch of 
our method is available in~\cite{GlazekLC2016}. 
  
\subsection{ Renormalized heavy-flavor Hamiltonian }

Starting from the one-flavor QCD Lagrangian~\cite{PDG2016},
\begin{eqnarray}
\cL\es \bar \psi (i\slashed D - m)\psi - \h \text{tr} F^{\mu\nu}F_{\mu\nu} \ ,
\end{eqnarray}
working in the gauge $A^+=0$ and using standard notation
for all tensors exemplified by $x^\pm = x^0 \pm x^3$, $ x^\perp = (x^1, x^2)$,
we calculate the canonical FF Hamiltonian 
\begin{eqnarray}
\hat H_{QCD}^{\rm can} \rs \hat P^-
\rs
\h \int dx^- d^2 x^\perp \  : \hat {\cal H}_{x^+=0} : \ ,
\label{Pm}
\end{eqnarray}
where $\cH$ is the well-known Hamiltonian density written in 
terms of quantum fields and their derivatives, see 
{\it e.g.}~\cite{Brodsky-Pauli-Pinsky}. 

The canonical Hamiltonian couples point-like particles with increasing 
strength when the invariant mass change it induces increases. 
This 
leads in perturbative calculations to divergent integrals over quark or 
gluon transverse momenta $k^\perp$, called UV divergences, and to 
diverging integrals over their $k^+$ momenta near zero, called small-$x$ 
divergences, since the ratio $x = k^+/P^+$, where $P^+$ denotes a 
hadron momentum, corresponds to the momentum fraction $x$ carried 
by a quark or gluon in the parton model. Therefore, our starting Hamiltonian 
includes the regularization described in~\ref{regular}. It also includes 
the corresponding counterterms.

After regularization, we introduce the effective particles of size 
$s$ by means of a unitary transformation of field operators, 
\beq
\label{qs}
\psi_s \es \cU_s \, \psi_0 \, \cU_s^\dagger \ ,
\eeq
and similarly for the gluon field $A$. The subscript 0 refers to the 
size $s=0$. The Hamiltonian is to be independent of the scale $s$,
\beq
\label{cHt}
H_s(q_s) \es H_0(q_0) \ ,
\eeq
where $q$ denotes quark and gluon operators. This condition is 
satisfied if $\cH_s = H_s(q_0)$ solves the RGPEP differential equation with respect to parameter 
$t = s^4$, introduced for convenience in handling dimensionful
quantities. We use the subscript $t$ below as equivalent to $s$. 
The RGPEP equation reads 
\beq 
\cH'_t \es
[ \cG_t , \cH_t ] \ ,
\label{RGPEP}
\eeq 
where $\cG_t = - \cU_t^\dagger \cU'_t$ is called the RGPEP 
generator. The initial condition at $t=0$ is set by the regulated 
canonical Hamiltonian with counterterms. The new generator
we use is~\cite{pRGPEP}
\beq
\label{cG2}
\cG_t \es [ \cH_f, \tilde \cH_t ] \ ,
\eeq
where the operator $\cH_f$ is the free part of $\cH_t$ and the operator 
$\tilde \cH_t$ is directly related to $\cH_t$ as described in~\ref{AppRGPEP}. 
The RGPEP Eq.~(\ref{RGPEP}) is designed according to the 
principles of similarity renormalization group procedure~\cite{SRG1,SRG2}, 
taking advantage of features of the double-commutator flow 
equations~\cite{Wegner1,Wegner2,Kehrein,dc0,dc1,dc2,dc3}. 

We solve Eq.~(\ref{RGPEP}) expanding $H_t(q_t)$ in powers of 
the effective coupling constant, which is assumed small on the 
basis of asymptotic freedom in our approach~\cite{AF}.
Namely, we take advantage of the hierarchy of scales of the heavy quark 
mass $m$, inverse effective particle size $s^{-1}$ and 
$\Lambda_{QCD}$ in our scheme,
\begin{eqnarray}
m \gtrsim  s^{-1} \gg \Lambda_{QCD} \ .
\label{scales}
\end{eqnarray}

\subsection{ Bound state problem for heavy quarkonia }

Quarkonium is defined as a solution to the eigenvalue problem 
of the renormalized Hamiltonian, $H_t(q_t)$, denoted for brevity 
by $H_t$, for eigenvectors with corresponding quantum numbers,
\begin{eqnarray}
H_t | \Psi \rangle 
\es 
 E | \Psi \rangle \ .
\label{eigenproblemHt}
\end{eqnarray}
The state $|\Psi\rangle$ is written in terms of the Fock components 
built using creation operators for the corresponding effective particles,
\beq
| \Psi \rangle 
\es
| Q_t \bar Q_t \rangle 
+ 
| Q_t \bar Q_t G_t \rangle 
+
| Q_t \bar Q_t G_t G_t \rangle 
+ ... \ .
\label{statePsi}
\eeq
In principle, this representation has infinitely many terms.
However, 
the interaction Hamiltonian that solves the RGPEP 
Eq.~(\ref{RGPEP}) is characterized by the vertex form factors, $f_{c.a} = e^{- t \ \left( \cM_c^2 - \cM_a^2 \right)^2 }$, 
where $\cM_c$ and $\cM_a$ denote the invariant masses of 
particles created and annihilated in the vertex, respectively.
The greater $t$ the stronger suppression of interactions that
change the number of virtual particles, and one may hope 
that the expansion may converge, or even a few terms may
be sufficient to obtain a reasonable first approximation. However,
the increase of size $s$, or parameter $t$, is associated with
the increase of the coupling constant in $H_t$, so that one has to 
study the effective dynamics in order to determine the values 
of $t$ that can be reliably achieved using an expansion of $H_t$ 
in a power series in $g$,
before one solves 
the resulting non-perturbative eigenvalue problem for $H_t$ 
expecting a desired accuracy. 
This 
approach to non-perturbative bound-state eigenvalue problems 
is supported by numerical studies of simple models with 
asymptotic freedom~\cite{GlazekWilson1998,GlazekMlynikAF}.

Using the RGPEP solution of the form,
\beq
\label{cHe}
H_t \es H_f + g H_{ 1 t} + g^2 H_{2 t} + ... \ ,
\label{Htexpanded}
\eeq
and keeping only terms up to second order in powers of the 
small coupling constant $g$, Eq.~(\ref{eigenproblemHt}) can 
be written as
\begin{align}
\left\{
\left[ \begin{array}{lll}
       . & . & . \\  
       . & H_f + g^2 H_{2t} &  g H_{1t}  \\  
       . & g H_{1t} & H_f + g^2 H_{2t}
       \end{array} \right]
-E 
\right\}
\left[  \begin{array}{l}
        .                    \\ 
        | Q_t \bar Q_t G_t \rangle   \\
        | Q_t \bar Q_t \rangle 
        \end{array} \right]
\rs 0 \, ,  
\label{eigenproblemmatrix}
\end{align}
where the dots represent the Fock components with more than 
one effective gluon and the Hamiltonian terms that involve
these components. The operator $H_{2t}$ is limited in 
Eq.~(\ref{eigenproblemmatrix}) to its terms that do not change the 
number of effective particles.

\subsection{ Gluon mass ansatz and effective eigenvalue problem }

Introducing the gluon mass term, denoted by $\mu^2$, as a minimal 
price one has to pay for dropping all the infinitely many dotted terms 
in Eq.~(\ref{eigenproblemmatrix}), one arrives at an approximate 
eigenvalue problem of the form
\beq
\left\{
\left[ \begin{array}{ll}
       H_f + \mu^2  &  g H_{1t}  \\  
       g H_{1t} & H_f + g^2 H_{2t}
       \end{array} \right]
- E 
\right\}
\left[  \begin{array}{l} 
        | Q_t \bar Q_t G_t \rangle   \\
        | Q_t \bar Q_t \rangle 
        \end{array} \right] 
\rs 0 \  .  
\label{meson1}
\eeq
 We also dropped the interaction terms $H_{2t}$ 
in the sector with one effective gluon because 
in the calculation that follows they contribute to 
the effective Hamiltonian for the $Q_t \bar Q_t$ 
component first in order $g^4$, while our calculation 
is limited here to terms order 1 and $g^2$.

The effective Hamiltonian for the $Q_t \bar Q_t$ component 
is calculated using the projection operator on that Fock sector, 
$\cP_t$, and the projector  $\cQ_t = 1 - \cP_t $ on the 
$Q_t \bar Q_t G_t$ sector.
\begin{align}
& |\Psi_\cP\rangle \equiv  \cP_t | \Psi \rangle  = | Q_t\bar Q_t \rangle \ ,
& |\Psi_\cQ\rangle \equiv  \cQ_t | \Psi \rangle = | Q_t\bar Q_t \, G_t\rangle  
\ .
\end{align}
Following Ref.~\cite{Wilson1970}, the component $|\Psi_\cQ\rangle$ 
is related to the component $|\Psi_\cP\rangle$ by the operator $\cR_t$,  
$|\Psi_\cQ\rangle  = \cR_t \, |\Psi_\cP\rangle$. 
The eigenvalue equation to solve is written as
\begin{eqnarray}
  H_{{\rm eff} \, t} | \psi_{Q\bar Q\,t} \rangle  \es E | \psi_{Q\bar Q\,t} \rangle  \ ,
\end{eqnarray}
where $|\psi_{Q\bar Q \, t}\rangle= S_t | Q_t\bar Q_t \rangle$,  
$S_t = (\cP_t+\cR_t^\dagger \cR_t)^{-1/2}$ and
\beq
 H_{{\rm eff} \, t} 
 \es
 S_t (\cP_t+\cR_t^\dagger)H_{t} (\cP_t+\cR_t) S_t \ .  
 \label{Heff}
\eeq
This formula is used below to evaluate the quarkonium 
effective Hamiltonian  $H_{{\rm eff} \, t}$ keeping terms
order 1 and $g^2$ and neglecting terms order $g^4$ 
and smaller. 

\section{ $Q\bar Q$ effective eigenvalue equation }
\label{secEVeq}

\begin{figure}
    \centering
 \includegraphics[width=6cm,height=1.7cm]{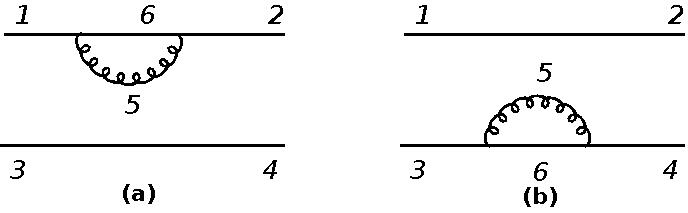}
    \caption{ Effective quark mass squared contributions in Eq.(\ref{BS-eq-summary}).}
    \label{massterms}
\end{figure}

\begin{figure}
    \centering 
    \includegraphics[width=8.2cm,height=1.7cm]{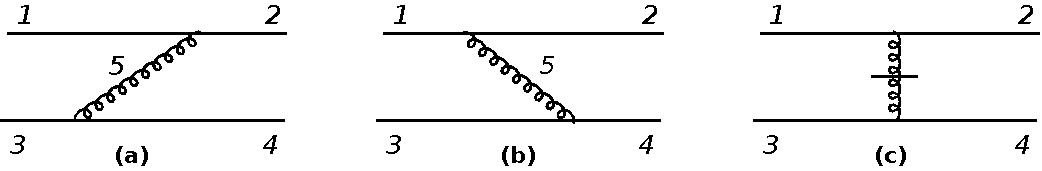}
    \caption{ Gluon-exchange and instantaneous interactions in
                 $ U_{t\,\rm{eff}} = H_{\rm exch}  + H_{\rm inst}$
                 of Eq.(\ref{BS-eq-summary}):
                 (a) corresponds to Eq.~(\ref{ZZ}),
                 (b) to Eq.~(\ref{trZZ}), and
                 (c) to Eq.~(\ref{inst}). }
    \label{interaction}
\end{figure}

The effective FF Hamiltonian eigenvalue equation for the 
heavy quarkonium $ Q\bar Q $  component reads
\begin{eqnarray}
H_{t\,\rm{eff}}|\psi_{Q\bar Q \,t}\rangle
\rs
\frac{M^2 + P^{\perp 2}}{P^+} |\psi_{Q\bar Q \,t}\rangle \ ,
\end{eqnarray}
where $P$ is the quarkonium kinematical momentum. Its 
mass is denoted by $M = 2m + B$, where $B$ is traditionally 
called the binding energy, although in QCD it may be positive. 
The kinematical momentum drops out of the eigenvalue equation 
for the quarkonium mass squared, 
\begin{eqnarray}
( P^+\, H_{t\,\rm{eff}}-P^{\perp 2}-M^2 ) |\psi_{Q\bar Q \,t}\rangle
\rs 0 \ . 
\end{eqnarray}
Using the two-body relative motion wave function $\psi_{t 24}(\kappa_{24}^\perp,x_2)$
for effective quarks of size $s = t^{1/4}$ and momenta $k_2$ and $k_4$,
where $x_2 = k_2^+/P^+$ and $\kappa_{24}^\perp = x_4 k^\perp_2 - x_2 k^\perp_4$,
the eigenstate is written as 
\begin{align}
&|\psi_{Q\bar Q t}\rangle
=
\sum_{~~~~~ 24} \hspace{-15pt}\int  P^+ 
\frac{\delta_{c_{2}c_{4}} }{ \sqrt{3}}  \tilde\delta (P-k_2-k_4) \, 
\psi_{t\,24}(\kappa^\perp_{24},x_2)  b_{2t}^\dagger d_{4t}^\dagger |0\rangle  .
\label{wf}
\end{align}
The sum extends over quarks spins and colors, integrations over kinematical
momenta are carried out with Lorentz-invariant FF measure $[k] = dk^+ d^2k^\perp/
[2 k^+ (2\pi)^3]$, factor $1/\sqrt{3}$ takes care of normalization of a color singlet 
described by the Kronecker delta in the quark color indices, and the tilde indicates 
that the kinematical total momentum conservation $\delta$-function is multiplied by 
$16\pi^3$. The mass-squared eigenvalue equation reads
\begin{align}
&
\left( 
\frac{\kappa_{13}^{\perp 2} + {\mathscr M}^2_{1,\,t}}{ x_1 } 
+
\frac{\kappa_{13}^{\perp 2} + {\mathscr M}^2_{3,\,t} }{ x_3 } 
- M^2 \right)\, \psi_{t\,13}(\kappa^\perp_{13},x_1) 
\nn 
&+
g^2 \int \frac{dx_2 d^2 \kappa_{24}^{\perp}}{ 2 (2\pi)^3 x_2 x_4} \, U_{t\,\rm{eff}}(13,24)
\, \psi_{t\,24}(\kappa^\perp_{24},x_{2})
\ = \ 0 \ , 
\label{BS-eq-summary}
\end{align}
where ${\mathscr M}^2_{i,\,t}$ is the effective mass for the quark, 
$i=1$, or antiquark, $i=3$. It contains the quark mass parameter
$m^2$ appearing in $H_f$ and the self-interaction contributions
that include UV-finite parts of the mass counterterm in the initial
QCD canonical Hamiltonian, $g^2 m_\delta^2$. Namely, as indicated
in Fig.~\ref{massterms},
\begin{align}
\mathscr M_{1,\,t}^2 
&= 
m^2 + g^2 \, m^2_\delta + \frac{4}{3} 
g^2 \int \frac{ d^2 \kappa^{\perp } dx }{ 2(2\pi)^3 x ( 1 - x ) }
\, r_{65.1}^2  \,
\nn 
&\times 
\left[ 
\sum_6 (\bar u_{1} \gamma^\mu  u_6) 
\, (\bar u_6 \gamma_\mu u_{1})
+ ( m^2 - \cM^2 ) 
\, 4 \frac{ 1 - x }{ x }
\right] 
\nn 
&\times 
\left(
\frac{ 1 }{ m^2 - \cM^2 } 
 - \frac{ 1 }{ m^2 - \cM^2 - {\mu^2_{653} / x} }
\right) \rm{e}^{-2\left( \cM^2-m^2 \right)^2t}
\ ,
\label{Mti}
\end{align}
and analogously for $\mathscr M_{3,\,t}^2 $. 
The gluon mass appears  with subscripts, such as $\mu^2_{ 653}$ above, because the ansatz $\mu^2$ in Eq.~(\ref{meson1}) is {\it a priori} allowed to be a function of the relative motion of a gluon with respect to quark and antiquark in the meson $Q_t \bar Q_t G_t$ component and the subscripts indicate the  three-particle state in which the ansatz function is evaluated, the gluon being always labeled by 5 as in Figs.~\ref{massterms} and \ref{interaction}. 
$\cM^2$ 
denotes an invariant mass of the fermion and boson in the loop with relative momenta 
$x\equiv x_{5/1}={k^+_5/k^+_1}$ and
$\kappa^\perp = (1-x) \, k_5^\perp - x \,k_6^\perp$.

The interaction kernel $U_{t\,\rm{eff}}(13,24)$ contains 
instantaneous FF interactions and gluon-exchange terms. There are 
two kinds of exchange terms. One kind comes directly from the
Hamiltonian $H_t$ in Eq.~(\ref{cHe}), which is obtained from canonical 
one-flavor QCD with massless gluons. This kind leads to terms denoted 
below by $F_{Z}$ and $F_{\trZ}$ that contribute to the Coulomb 
potential with Breit-Fermi spin-dependent factors. The other kind of 
terms comes from the exchange of the effective gluon, for which we have 
introduced the mass ansatz. This other kind leads to the terms denoted 
below by $R_{Z}$ and $R_{\trZ}$. These terms contribute the 
spin-independent oscillator correction to the Coulomb interaction.

In the notation of Fig.~\ref{interaction}, $ U_{t\,\rm{eff}}
=
H_{\rm exch}  + H_{\rm inst}$ and
\begin{align}
H_{\rm exch } 
&=
- \, \frac{4}{3} \, \left[
\, \frac{ \theta(x_1-x_2) }{ k^+_5} \, 
r_{25.1} r_{35.4} \,
d_{\mu\nu} (k_5) ( \bar u_1 \gamma^\mu u_2) (\bar v_4 \gamma^\nu v_3) 
\, \cF_{\trZ}
\right. \nn
& + 
\left.
\frac{ \theta(x_2-x_1) }{ k^+_5} \, 
r_{15.2} r_{45.3} \,
 \,
d_{\mu\nu} (k_5) (\bar v_4 \gamma^\mu v_3) ( \bar u_1 \gamma^\nu u_2)
\, \cF_{Z} 
\right]   , \\
H_{\rm inst}  &= 
- \, 
r_{C 13.42} \, f_{t\, 13.24 } 
\, 4 \, \sqrt{x_1 x_2 x_3 x_4} \,
  \frac{ \delta_{12} \delta_{34} }{ (x_1-x_2)^2}\,
\frac{4}{3} \ ,
\label{inst}
\end{align}
where $d_{\mu\nu}(k_5)
=
- g_{\mu\nu}
+ (n_\mu k_{5\nu} + n_\nu k_{5\mu}) / k_5^+ $ and
\begin{eqnarray}
\label{trZZ}
\cF_{\trZ} =   
f_{t\, 13.24 } \, F_{\trZ}
+ f_{t\, 4.53 } \, f_{t\, 1.52 } \, R_{\trZ} \ , \\
\label{ZZ}
\cF_{Z} =
  f_{t\, 13.24 } \, F_Z
+ f_{t\, 3.54 } \, f_{t\, 2.51 } \, R_{Z}
\ , 
\end{eqnarray}  
with
\begin{align}
F_{\trZ}  
& \rs
- 
\left( f_{t\, 1.52 } f_{t\, 4.53 }  - f_{t\, 13.24 } \right) f^{-1}_{t\, 13.24 }
\nn
& \times
\frac{ k_1^+(m^2 - \cM_{25}^2) + k_4^+(m^2-\cM_{35}^2)
}{
(m^2 - \cM_{25}^2)^2 + (m^2-\cM_{35}^2)^2 - (\cM_{13}^2 - \cM_{24}^2)^2 } \ , 
\label{Ftrz} \\
R_{\trZ}
& \rs
\h
\left(\frac{ k^+_1 }{  m^2 - \cM^2_{25} - 
\frac{x_1}{x_5} \mu^2_{253} }
+
\frac{k^+_4 }{ m^2 - \cM^2_{53} - 
\frac{x_4 }{ x_5} \mu^2_{253} }  \right) \ , 
\end{align}
and
\begin{align}
F_{Z} 
& \rs
- 
\left( f_{t\, 2.51 } f_{t\, 3.54 } - f_{t\, 13.24 } \right) f^{-1}_{t\, 13.24 }
\nn 
&\times 
\frac{ k_3^+(m^2 - \cM_{45}^2) + k_2^+(m^2 - \cM_{15}^2)
}{
 (m^2 - \cM_{45}^2)^2 + (m^2 - \cM_{15}^2)^2 - (\cM_{13}^2 - \cM_{24}^2)^2}  \ , 
 \label{Fz}\\
R_{Z}
& \rs
\h 
\left( \frac{k^+_2 }{  m^2 - \cM^2_{15} - 
\frac{x_2}{x_5} \mu^2_{154} } +
\frac{ k^+_3 }{  m^2 - \cM^2_{54} - 
\frac{ x_3 }{ x_5 } \mu^2_{154} }
\right) \ . 
\end{align}
Symbol $\mu^2_{i5j}$ denotes the value of the
gluon mass ansatz function for the intermediate
state of quark labeled by $i$, gluon by $5$ and
antiquark by $j$.
The factors $r$  with subscripts indicating particle momenta are the 
regulating functions in canonical QCD, see \ref{regular}.
Factors $f$ and $F$ result from solving Eq.~(\ref{RGPEP}). 
The factor $F$ differs from the analogous one in~\cite{QQ1}
due to the new generator. The difference is the appearance
of an additional term, $-(\cM_{13}^2 - \cM_{24}^2)^2$,
in the denominators and an additional $f_{t\,13.24}$ dividing
other two $f$s in Eqs. (\ref{Ftrz}) and (\ref{Fz}).  The subscripts Z 
and \rZ correspond to the gluon exchange as shown in Fig. \ref{interaction}.
Factors $R$ come from the operator $\cR_t$ in second order
terms in Eq.~(\ref{Heff}). They contain only the first-order
RGPEP form factors, which are the same as for the old generator,
used in~\cite{QQ1}. The subscripts 
of the form factors $f_{t\,a.b}$ indicate the dependence on 
the momentum variables.
For example, 
$f_{13.24} = \exp{[ - ( \cM_{13}^2 - \cM_{24}^2 )^2 t]}$, 
$f_{2.51}   = \exp{[- ( \cM_{51}^2 - m^2 )^2 t]}$ and
$f_{3.54}   = \exp{[- ( \cM_{54}^2 - m^2 )^2 t]}$.

\subsection{Small-x and heavy-quark limits with the new generator}

The new term $-(\cM_{13}^2 - \cM_{24}^2)^2$ appearing in 
the form factor $f_{t \, 13.24}$ and denominators of Eqs.~(\ref{Ftrz}) and 
(\ref{Fz}) due to the new RGPEP generator, does not matter in the leading 
behavior of functions $F$ when $x\to 0$. Thus, our analysis of the small-$x$ 
dynamics in the effective $Q_t \bar Q_t$ sector obtained using the new 
RGPEP generator leads to the same findings as in~\cite{QQ1}. If in the 
limit $x_5 \to 0 $ the ansatz $\mu^2$ in the $Q_t \bar Q_t G_t$ sector 
behaves like $\sim \kappa_5^{\perp 2} x_5^{\delta_\mu}$ with $ 0 < 
\delta_\mu <1/2 $ , then all small-$x$ divergences cancel out. 
The variables $x_5$ and $\kappa_5^\perp$ denote the relative
momenta of the gluon with respect to quark-antiquark pair in the $Q_t \bar Q_t G_t$ sector, {\it cf.} Figs.~\ref{massterms}
and \ref{interaction}. 

The heavy-quark limit of the $Q_t \bar Q_t$ 
interaction is also not affected by the generator change because, in 
comparison to the other terms that are the same as with the old 
generator, the new term is suppressed by the square of the ratio of 
relative quark-antiquark momentum to their mass. These are the two 
bottom-line reasons for the two different RGPEP generators to produce 
the same effective $Q_t \bar Q_t$ dynamics.

\section{ Harmonic oscillator force }
\label{secHO}

The scale hierarchy of Eq.~(\ref{scales}) allows one to consider the 
eigenvalue equation in the non-relativistic approximation for relative
motion of quarks, because the RGPEP factors exclude invariant mass 
changes greater than $s^{-1}$ and the dominant relative momenta 
of quarks are smaller than the quark mass. We introduce the quark 
relative momenta $\vec k_{ij} =(k_{ij}^\perp, k_{ij}^3)$, noting that 
analogous variables appear in light-front holography~\cite{holography,TrawinskiConfinement},  
\begin{align}\label{kij}
k^\perp_{ij} 
&\rs \h \frac{ \kappa^\perp_{ij} }{ \sqrt{x_i x_j} } \ ,
& k_{ij}^3 
&
\rs 
\frac{ m }{ \sqrt{ x_i x_j} } \left( x_i - \h \right) \ .
\end{align}
and we define the momentum transfer $\vec q = \vec k_{13} - \vec k_{24}$.
The non-relativistic approximation is obtained in the limit $ \vec k_{ij} / m 
\to 0 $. The eigenvalue equation in this limit reads 
\begin{align}
& \left[
{ {\vec k}^{\, 2}_{13 } \over m} - B
+ {\delta m_{1\,t}^2 \over 2 m } 
+ {\delta m_{3\,t}^2 \over 2 m } \right]\, \psi_{13}(\vec k_{13}) \nn
& + \int {d^3 q 
\over (2\pi)^3 } \, 
\left[V_{C,BF} ( \vec q\, ) + W ( \vec q\, )\right]
\, \psi_{24}(\vec k_{13}-\vec q)
\rs 0 \ ,
\end{align}
where the Coulomb potential with Breit-Fermi spin-dependent
factors is
\begin{align}
V_{C,BF} ( \vec q\, )
& \rs
- {4 \over 3}  
{ 4 \pi \alpha \over \vec q \,^2 } 
\left( 1 \ + \ BF \right)  \ , 
\end{align}
with $\alpha = g^2/(4\pi)$, 
and the additional term contains
\begin{align}
W ( \vec q\, )
& \rs
{4 \over 3} \, 4 \pi \alpha
\left[ 
{ 1 \over \vec q\,^2 }
- { 1 \over q_z^2 }
\right]\, 
{ \mu^2 \over \mu^2 + \vec q\,^2  }
\, e^{- 2 t m^2 |\vec q\,|^4 / q_z^2 }
\ ,
\label{W}
\end{align}
with $\mu^2 
= \theta(q_z) \, \mu_{253}^2 + 
\theta(-q_z) \, \mu_{154}^2$.  
The quark mass corrections are given by the same function $W$,
$\delta m_{i\,t}^2/ m 
=
-\int [d^3 q/ (2\pi)^3 ] W (\vec q\,)$,
where  $\mu^2=\mu_{653}^2$ for $i=1$ and $\mu^2=\mu_{651}^2$
for $i=3$, and 
in the mass counterterm we have chosen $m_\delta^2=0$. 
The only dependence on the gluon mass ansatz
function $\mu$ appears in the factor $\mu^2/(\mu^2
+ q^2)$, which could be replaced by one in any
integral involving $W$ if the mass $\mu$ dominated
the momentum $q = |\vec q\,|$ in the relevant integration
range. It is shown below that such dominance appears
indeed in the heavy quark limit, see Eq.~(\ref{taunu}).
This implies that the 
mass ansatz allows us to finesse the universal oscillator result irrespective
of the ansatz actual value and the choice of the RGPEP generator.

Once one renames the momentum  $\vec k_{13}$ as $\vec k $ 
and suppresses spin subscripts that are not important for the 
spin-independent correction to the Coulomb term, one obtains 
the same result as in Eq.~(105) of \cite{QQ1} despite that we use 
the new generator and holography-motivated relative momentum 
variables,
\begin{align}
&\left[ { \vec k^2 \over m} - B\right] \psi (\vec k)
+ \int { d^3 q \over (2\pi)^3} V_{C,\, BF} ( \vec q )
\, \psi (\vec k - \vec q )
\nn 
& + \int { d^3 q \over (2\pi)^3} W (\vec q) 
\left[ \psi (\vec k - \vec q )- \psi (\vec k) \right] 
= 0
\ .
\end{align}
Since the RGPEP form factors limit $\vec q$ to small values 
when $q_z$ is small, one can expand the wave function
in the small region of integration over $\vec q$ around 
$\vec k$, 
\begin{eqnarray}
\psi( \vec k - \vec q\, ) 
=
\psi( \vec k ) 
- q_i { \partial \over \partial k_i } \psi( \vec k )
+ \h q_i q_j { \partial^2 \over \partial k_i \partial k_j } \psi( \vec k ) + \dots
\end{eqnarray}
and observe that only even terms contribute, because $W$ 
is an even function of $\vec q$. The quadratic terms dominate 
and yield the harmonic oscillator potential, which is thus shown 
to not depend on the change in the generator,
\begin{align}
& - { 4 \over 3 } { \alpha \over 2\pi } b^{-3} \sum_i \tau_i { \partial^2 \over \partial k_i^2 } \ , 
\end{align} 
with 
$b = \sqrt{2m^2t} $, the vector 
 $\vec \tau  = \int_{0}^{1} dv \, v ( 1 - v^2 ) \, \vec w (v) \, \tau(v) $,
and $v = q_z/|\vec q\,|$ in $\vec w = (1-v^2, 1-v^2, 2 v^2)$.
The function $\tau(v)$,
\begin{align}\label{taunu}
  \tau(v)
  & = 
\int_0^\infty du \
u^2 \ e^{- u^2 } \ 
\left[ 1  + { u^2 \ v^2 \over 2 \ (m \, s)^2 \ (\mu \, s)^2 }  \right]^{-1}
\ ,
\end{align}
results from the first-order RGPEP form factor 
and the gluon mass ansatz. The form factor is the same for 
the new generator as it was for the old one.
It is visible that in the limit of heavy quarks
the function $\tau(v)$ is a constant  $\sqrt{\pi}/4$
no matter how large is the gluon mass ansatz $\mu$,
and a constant $\tau$ yields a rotationally symmetric
harmonic oscillator potential.
The resulting 
effective Schr\"odinger equation in momentum space reads
\begin{eqnarray}
\left[ { \vec k\,^2 \over m} - \h \, \tilde \kappa \, \Delta_{\vec k} - B\right] \psi (\vec k)
+ \int { d^3 q \over (2\pi)^3} V_{C,\, BF} ( \vec q )
\, \psi (\vec k - \vec q )
= 0 \ ,
\label{EqMomentumSpace}
\end{eqnarray}
with $\tilde \kappa = m \omega^2/2 
= \alpha \ (m s^2)^{-3} /( 36 \sqrt{2\pi})$. 
In the heavy quark limit, the harmonic oscillator
frequency $\omega$ depends on the mass and
size of the effective quarks but not on the value
of the gluon mass ansatz. This result is formally
valid as long as the gluon mass ansatz is not
zero.
When the gluon mass ansatz is set to zero, the harmonic 
oscillator potential vanishes and the Coulomb term is the 
same as in QED, except for the color factor 4/3 and the strong
coupling constant $g$ replacing the electric charge $e$. 

Although Eq.~(\ref{EqMomentumSpace}) with Breit-Fermi
terms in new variables of Eq.~(\ref{kij}) has not
been solved yet, estimates previously attempted~\cite{QQ2}
using the old version of the RGPEP and standard FF variables
suggest that the values of size $s$ and frequency $\omega$
that could be useful in the phenomenology of charmonium and
bottomonium families using our new RGPEP version, may be on
the order of $1/m$ and $200-300$ MeV, respectively. For such
values, according to Eq.~(\ref{taunu}), a rotationally symmetric
eigenvalue problem in the new variables may emerge if the gluon
mass ansatz is either very large in comparison of the quark
mass or it is some function of the gluon motion with respect
to quarks that leads to $\vec \tau$ with three equal components,
such as, for example, $\mu^2 \sim v^2$. The first possibility
appears unlikely when one assumes that the gluon mass should
be on the order of at most few GeV for a realistic value of
the coupling constant. The second option requires insight
concerning the functional form implied by QCD. Therefore,
the most desired course of study is to estimate the gluon
mass terms in QCD using the 4th order RGPEP.

\section{Conclusion }
\label{secConclusions}

Since the effective $Q \bar Q$ Hamiltonian we obtain using the RGPEP is 
the same for different generators and does not depend on the gluon
mass ansatz value, one may hope that it may serve as a reasonable initial 
approximation to one heavy-flavor QCD in the range of momentum scales 
at which the mechanism of binding of effective quarks by effective gluons 
is at work. The question 
arises if a more accurate calculation, using the
solution to 
Eq.~(\ref{RGPEP}) with accuracy to terms order 
$g^4$ and shifting the gluon mass ansatz to the two-gluon component, 
also leads to the same oscillator potential. An additional reason for interest 
in such calculation, 
not addressed here,
stems from the fact 
that the eigenvalue problem derived in a similar way for a single quark, 
yields an infinite eigenvalue, indicating how the renormalized effective 
particle approach can tackle the issue of confinement. 

Our assumption that the gluon mass ansatz can mimic non-perturbative
non-Abelian dynamics in higher Fock components is testable in two ways. 
One way is the suggested above RGPEP calculation of higher order than 
$g^2$ combined with a shift of the mass ansatz to components with more 
effective gluons than one. The other way is to include non-perturbative 
running of the constituent gluon and quark masses with their size parameter 
$s$, which can be approximately described using Eq.~(\ref{RGPEP}).

Two qualitative arguments support our expectation that a harmonic oscillator 
interaction will also result from more advanced RGPEP calculations than carried
out here. The first argument is that the eigenvalue of the effective FF Hamiltonian 
is the square of the quarkonium mass. In contrast, in the instant form (IF) of 
dynamics the eigenvalue is the quarkonium mass itself. Therefore, if the 
quark-antiquark potential commonly used to describe confinement 
in the IF Hamiltonians is linear in the distance between quark and antiquark, then 
the corresponding FF Hamiltonian should be quadratic~\cite{TrawinskiConfinement}. 
This correspondence between linear and quadratic potentials for heavy quarkonia
needs to be verified numerically. The second argument is that the type of 
momentum variables that we use and the quadratic nature of effective potential 
both appear also in the light-front holography for light hadrons. The holography 
claims to represent the first approximation to QCD~\cite{holography}. This coincidence 
suggests the dynamical utlity of the concept of effective particles in QCD, 
corresponding to the constituent quarks in light hadrons or charm and bottom 
quarks in heavy hadrons, both kinds potentially describable by including effective 
gluons in the RGPEP.

Even the crude oscillator picture developed here is able to provide wave functions 
that can be applied in phenomenology of heavy quarkonia~\cite{MatsuiSatz1986,
Shan:2017uts,Andronic:2015wma,Rapp:2008tf,Aaij:2015awa,SzczurekLHC,
SzczurekJPsi,SzczurekPANDA,ChineseQuarkonia,Karliner} and that can be compared 
with results of other approaches~\cite{Pineda,BlankBottomonium,Hilger:2014nma,
PopoviciQuarks,FischerKubrak,Leitao:2016bqq,SegoviaArriola,GuoSzczepaniakHybrids,QhLebedSwanson,Riek:2010fk}, 
including those that use the FF of Hamiltonian dynamics~\cite{BrisudovaPerryWilson,BrisudovaMassiveGluons,AmesQQHolography,AmesQQLF}.
Numerical calculations are needed to determine the probability of $Q_s \bar Q_s G_s$ 
component. The probability of two-gluons component is needed to find out if the expansion 
in the number of effective gluons of appropriate mass and size has a chance to converge.
This requires fourth-order calculations.  Such
calculations are also needed for the RGPEP to
complement other approaches addressing
non-Abelian dynamical effects,
such as Dyson-Schwinger equations~\cite{Aguilar:2008xm,Binosi:2009qm,Aguilar:2009nf}, functional-renormalization group~\cite{AlkoferSmekal,
Pawlowski:2005xe,FischerMaasPawlowski}, and
other Hamiltonian methods~\cite{Popovici:2010mb,Heffner:2012sx,GreensiteSzczepaniak2016}.
If the RGPEP calculations 
did show convergence in the number of effective gluons, the effective quantum field 
operators it introduces might become useful in designing optimized interpolating operators 
for lattice studies~\cite{JLabOperators} and obtaining the Minkowski space images of 
hadrons. 

\black

\appendix

\section{ Regularization }
\label{regular}

We regulate the QCD interaction terms.  Let $k^\perp$ and $k^+$ label 
transverse and longitudinal momenta of a particle
taking part  in an 
interaction. The total kinematical momentum annihilated in a vertex is 
labeled by $P^\perp$ and $P^+$. The relative transverse momentum 
of any particle of momentum $k$ that is involved in the vertex, with respect 
to all other particles in that vertex, is defined by $\kappa^\perp = k^\perp - 
x P^\perp$, where $x=k^+/P^+$. Every creation or annihilation operator
for a quark in every interaction vertex is supplied with a vertex factor
$\exp{[-(m^2 + \kappa^{\perp \, 2})/(x \Delta^2)]} $,
and every gluon operator is supplied with a factor
$\exp{[-(\delta^2 + \kappa^{\perp \, 2})/(x \Delta^2)]}$.
The parameter $\Delta \to \infty$ when the ultraviolet regularization 
is being removed. The parameter $\delta$ plays the role of an 
infinitesimal gluon mass that regulates small-$x$ divergences
by the parameter $\epsilon = \delta/\Delta $.
The small-$x$ regularization is being removed by $\epsilon \to 0$.

For example, in an interaction vertex where a quark with
momentum $k_3$ splits into a quark with momentum $k_2$
and a gluon with momentum $k_1$, the regulating function
is
\begin{align}
r_{21.3}
\rs
e^{-\frac{m^2 + \kappa_{2/3}^2}{x_{2/3} \Delta^2}}
e^{-\frac{\delta^2 + \kappa_{1/3}^2}{x_{1/3} \Delta^2}}
e^{-\frac{m^2}{\Delta^2}}
\rs
e^{-\frac{\cM^2_{21\,\delta} + m^2}{\Delta^2}}
\ .
\label{reg213}
\end{align}
$\cM^2_{21\,\delta}$ denotes invariant mass of particles $k_1$ and $k_2$, 
in which the gluon has mass $\delta$,
$\cM^2_{21\,\delta} = \cM^2_{21} + \delta^2/x_{1/3}$.

The FF canonical Hamiltonian contains also interactions
with 
instantaneous exchange of a gluon or a quark. 
These
terms are singular as functions
of the exchanged longitudinal momentum. 
They are regulated as if they were made of two local 
interaction vertices. For example, the instantaneous interaction 
between a quark and an antiquark 
(with ingoing and outgoing 
momenta 
$k_2$, $k_4$ and $k_1$, $k_3$, respectively) is 
regulated as if they exchanged a gluon with momentum 
$k_5=(k_1-k_2)z/|z|$, where $z=x_1-x_2$. The regulating 
function in this case is
\beq
r_{C13.42}
\rs
  \theta(z) \ r_{25.1} r_{35.4}
+ \theta(-z)\ r_{45.3} r_{15.2}
\ .
\label{regC1342}
\eeq
%
\section{ Elements of RGPEP }\label{AppRGPEP}
The operator $\cH_f$, called the free Hamiltonian, is the kinetic term, 
which does not contain the QCD coupling constant $g$.
\beq
\label{cHf} 
\cH_f =
\sum_i \,  { p_i^{\perp \, 2} + m_i^2 \over p_i^+}  \, q^\dagger_{0i} q_{0i} \, ,
\eeq 
where $i$ denotes particle species.
The operator $\tilde \cH_t$ for any 
\beq
\label{Hstructure} 
\cH_t =
\sum_{n=2}^\infty \, 
\sum_{i_1, i_2, ..., i_n} \, c_t(i_1,...,i_n) \, \, q^\dagger_{0i_1}
\cdot \cdot \cdot q_{0i_n} \, ,
\eeq 
is defined by multiplication of each and every term in it 
by a square of a total $+$ momentum involved
in a term,
\beq
\label{HPstructure} 
\tilde \cH_t =
\sum_{n=2}^\infty \, 
\sum_{i_1, i_2, ..., i_n} \, c_t(i_1,...,i_n) \, 
\left( {1 \over
2}\sum_{k=1}^n p_{i_k}^+ \right)^2 \, \, q^\dagger_{0i_1}
\cdot \cdot \cdot q_{0i_n} \, .
\eeq 
The multiplication by this factor secures invariance of $H_s$ 
with respect to seven kinematical symmetries of the FF dynamics.

\bibliography{RGPEPrefs}{}
\bibliographystyle{elsarticle-num}

\end{document}